# Foundation model for particle collision data analysis

Ting-Hsiang Hsu (徐鼎翔)[1], Baihong Zhou (周柏宏)[2,3,4], Qibin Liu (刘齐斌)[5], Yue Xu (许悦)[6], Shu Li (李数)[2,3,4], George Wei-Shu Hou (侯維恕)[1], Benjamin Nachman[5,7], Shih-Chieh Hsu (徐士傑)[6], Vinicius Mikuni[8] ✉, Yuan-Tang Chou (周圓唐)[9] ✉ & Yulei Zhang (张宇雷)[6] ✉

Data-intensive collider experiments increasingly require models that transfer across high-dimensional, partially observed measurements, yet most analyses still optimize separate models for each task. Here, to address this bottleneck, we introduce EVENET, an event-level foundation model pretrained on 500 million simulated collision events using self-supervised learning and physics-informed supervision. The model constructs a shared particle-cloud representation from the variable-length set of physics objects in each event. Across searches for heavy resonances and exotic Higgs decays, EVENET outperforms both established and state-of-the-art baselines, particularly in low-statistics regimes. We validate transfer to experimental data by rediscovering the $\Upsilon$ meson in CMS Open Data. The pretrained model also supports precise extraction of quantum-correlation observables that remain stable under the tested systematic variations. Together, these results establish a reusable backbone for diverse collider analyses, improving downstream adaptation after a substantial one-time pretraining investment.

Foundation models have reshaped how large-scale data are analysed across science and engineering. In natural language processing, self-supervised pretraining on massive text corpora produces general-purpose representations that transfer efficiently to diverse downstream tasks with minimal fine-tuning, as demonstrated by BERT[1] and GPT[2]. In computer vision, contrastive learning and masked autoencoders[3,4] have matched or exceeded fully supervised baselines by learning from unlabelled images alone. More recently, this pretrain-then-transfer paradigm has permeated the natural sciences: protein language models predict atomic-level structures from amino acid sequences[5], and self-supervised encoders extract robust imaging biomarkers from limited clinical cohorts[6]. A common lesson from these domains is that large-scale pretraining on domain data captures latent structure that generalizes far more efficiently than task-specific training, particularly when labelled data are scarce[7].

High-energy physics faces a data landscape similar to that of other fields. The Large Hadron Collider (LHC)[8] delivers billions of proton–proton collisions per year, each recorded as a variable-length set of final-state particles whose properties encode the physics of the underlying hard scattering. Extracting insight from these high-dimensional, partially observed records is central to the LHC programme, spanning searches for new resonances, measurements of rare decay channels and precision tests of the standard model (SM). Machine learning has become essential for jet tagging[9,10], event classification[11], anomaly detection[12,13] and fast simulation[14], consistently surpassing traditional multivariate approaches[15]. Yet the prevailing paradigm remains one of task-specific optimization: for each new analysis, a dedicated model is designed, trained on purpose-built labelled simulations and validated independently. This imposes increasing costs in expert effort, computing resources and turnaround time that will intensify as the High-Luminosity LHC (HL-LHC)[16] increases both data volumes and the demand for sensitivity to rare processes. The same challenge also constrains current particle physics experiments, such as Belle II[17], DUNE[18] and future collider experiments.

[1]Department of Physics, National Taiwan University, Taipei, Taiwan. [2]State Key Laboratory of Dark Matter Physics, Tsung-Dao Lee Institute, Shanghai Jiao Tong University, Shanghai, China. [3]State Key Laboratory of Dark Matter Physics, Institute of Nuclear and Particle Physics, School of Physics and Astronomy, Shanghai Jiao Tong University, Shanghai, China. [4]Key Laboratory for Particle Astrophysics and Cosmology (KLPPAC-MoE), Shanghai Key Laboratory for Particle Physics and Cosmology (SKLPPC), Shanghai, China. [5]Fundamental Physics Directorate, SLAC National Accelerator Laboratory, Menlo Park, CA, USA. [6]Department of Physics, University of Washington, Seattle, WA, USA. [7]Department of Particle Physics and Astrophysics, Stanford University, Stanford, CA, USA. [8]Kobayashi-Maskawa Institute, Nagoya University, Nagoya, Japan. [9]Department of Physics, National Tsing Hua University, Hsinchu, Taiwan. ✉e-mail: vmikuni@hepl.phys.nagoya-u.ac.jp; ytchou@phys.nthu.edu.tw; zyl94@uw.edu

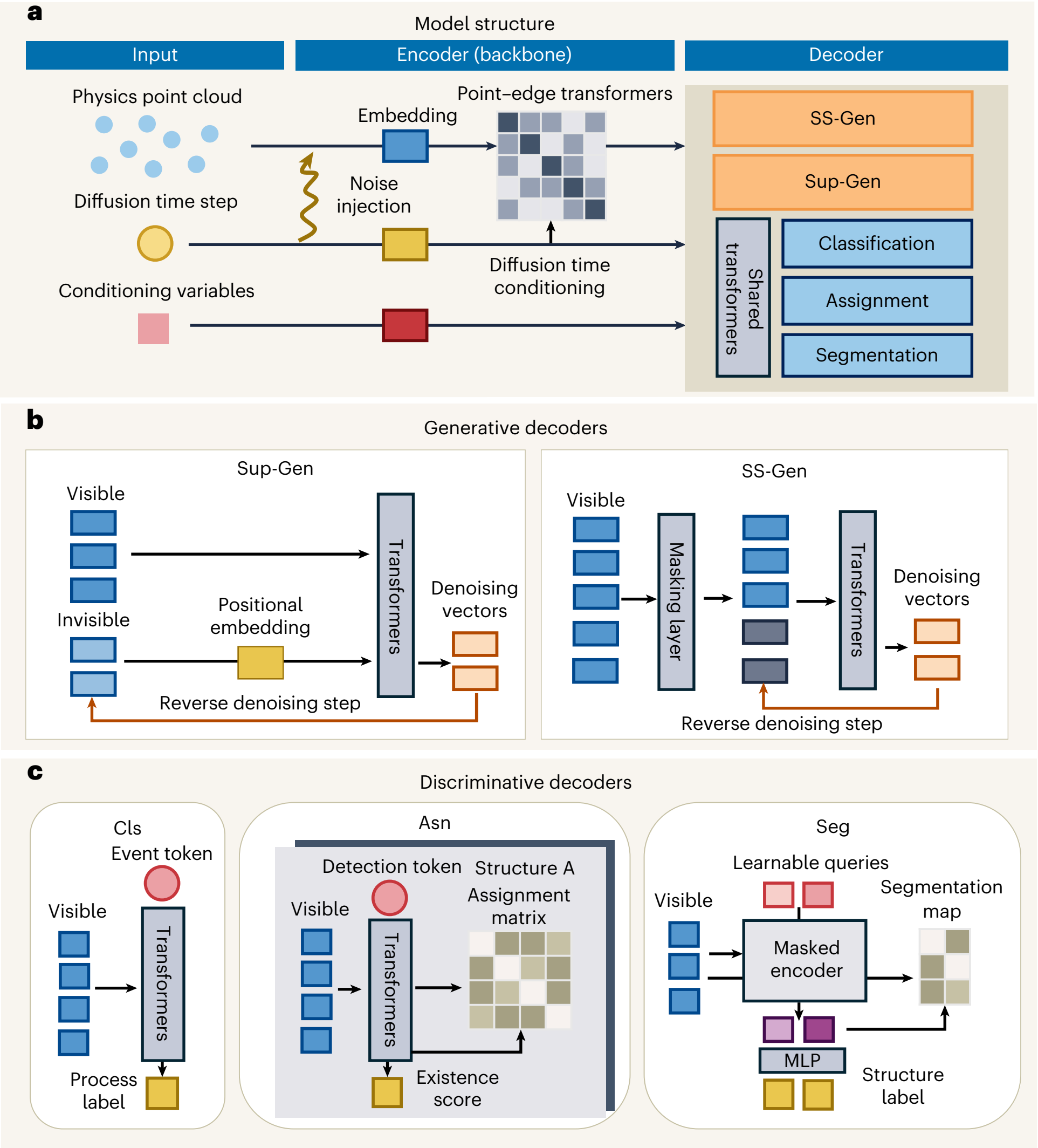

**Fig. 1 | Overview of the EVENET model architecture and tasks. a**, Collision events are reconstructed as sets of physics objects, which form the input to a point–edge transformer encoder. Task-specific decoders extend this shared representation to both generative and discriminative objectives. **b**, Self-supervised generation (SS-Gen) reconstructs masked input objects (or generates complete events), whereas supervised generation (Sup-Gen) predicts target particles conditioned on observed objects using diffusion. **c**, Discriminative decoders branch into three task heads aligned with common analysis goals: identifying the physics process (Cls), matching reconstructed objects to their parent resonances (Asn) and grouping objects into final-state categories for event interpretation (Seg). MLP, multilayer perceptron.

Initial steps towards foundation models in collider physics have focused on individual jets, learning substructure representations for tagging and classification[19–21]. Weakly supervised and self-supervised methods have also been explored for model-agnostic anomaly detection[22–27]. Although promising, these efforts remain limited in scope. Most operate on single physics objects rather than full collision events. Existing approaches for full events are typically limited to a few downstream tasks[28] or narrow physics process coverage[29], leaving open whether the learned representations transfer across diverse collider analyses. Furthermore, the critical step of validating learned representations on real experimental data, as opposed to simulation alone, has received little attention[30].

Here we introduce EVENET, an event-level foundation model for particle collision data that addresses these limitations (Fig. 1). EVENET represents each collision as an unordered particle cloud and processes it using a shared point–edge transformer backbone. Interchangeable task heads support process classification, object-to-resonance assignment, final-state segmentation, and diffusion-based reconstruction of masked or undetected particles. The backbone is pretrained on 500 million simulated events through a two-stage strategy combining self-supervised learning (SSL) with physics-informed supervision. We evaluate its generalization, simulation-to-data transfer, data efficiency, systematic robustness across new-physics searches, resonance rediscovery in open data from the Compact Muon Solenoid (CMS) experiment, and precision quantum-correlation measurements. The resulting backbone is designed for repeated downstream adaptation after a substantial one-time pretraining investment.

**Table 1 | Pretraining and downstream training configurations**

| Category | Cls | Seg | Asn | SS-Gen | Sup-Gen | Base LR | Optimizer | GPUs | Size (million) |
|---|---|---|---|---|---|---|---|---|---|
| **Pretraining stages** | | | | | | | | | |
| Stage I (SSL) | ✗ | ✗ | ✗ | ✓ | ✗ | $1.8\times10^{-5}$ | LION | 512 | 18.8 + 1.3 |
| Stage II (full) | ✓ | ✗ | ✗ | ✓ | ✓ | $1.2\times10^{-5}$ | LION | 512 | 18.8 + 3.8 |
| **Downstream tasks** | | | | | | | | | |
| Task 1 (Ind.) | ✓ | ✗ | ✗ | ✗ | ✗ | $3.0\times10^{-4}$ | AdamW | 1 | 6.3 + 1.2 |
| Task 1 (Param.) | ✓ | ✗ | ✗ | ✗ | ✗ | $1.0\times10^{-5}$ | AdamW | 2 | 18.8 + 1.2 |
| Task 2 | ✗ | ✗ | ✗ | ✓ | ✗ | $5.0\times10^{-4}$ | AdamW | 4 | 18.8 + 1.7 |
| Task 3 | ✓ | ✓ | ✓ | ✗ | ✗ | $4.0\times10^{-4}$ | AdamW | 4 | 18.8 + 8.7 |
| Task 4 | ✗ | ✗ | ✓ | ✗ | ✓ | $5.0\times10^{-5}$ | AdamW | 16 | 18.8 + 6.8 |

A tick indicates that the corresponding network head was used in the training. Base LR is the learning rate before square-root scaling with world size ($\mathrm{LR_{eff}} = \mathrm{Base\ \ LR} \times \sqrt{N_{\mathrm{GPU}}}$). Here $\mathrm{LR_{eff}}$ is the effective learning rate after scaling, and $N_{\mathrm{GPU}}$ is the number of GPUs used for distributed training. Size (million) reports trainable encoder and decoder parameters (in millions). Task 1: generalization across hypothesis grids for discovery; Ind. denotes per-hypothesis training; Param. denotes joint training across hypotheses. Task 2: simulation-to-data transfer with resonance rediscovery via particle-cloud generation. Task 3: new-physics search using exotic Higgs decays. Task 4: quantum-correlation measurement in dileptonic top-quark pair events.

# Results

We evaluate EVENET, an end-to-end event-level foundation model designed to learn a unified representation across a heterogeneous set of analysis tasks. EVENET operates directly on variable-length particle-cloud inputs using a single backbone that can be applied across diverse observational regimes, including fully reconstructed, partially reconstructed and noise-perturbed events.

The backbone is pretrained with approximately balanced event counts across the process–final-state channels listed in Extended Data Table 1, without cross-section-based event weighting or sampling, thereby avoiding the underrepresentation of rare processes. Pretraining follows the hybrid pretraining strategy described in 'Pretraining strategy selection': (1) a self-supervised masking phase, followed by (2) a joint optimization phase combining self-supervised and supervised objectives. We then benchmark a suite of downstream tasks that separately probe generalization ('Generalization across hypothesis grids for discovery'), simulation-to-data transfer ('Simulation-to-data transfer with resonance rediscovery'), multi-task reconstruction and robustness to detector-level systematic variations ('Physics-structured multi-task synergy and robustness') and training efficiency ('Training convergence').

## Pretraining strategy selection

To quantify the contribution of the proposed pretraining strategy, we compare three EVENET variants that differ only in their pretraining objectives while sharing an identical backbone architecture and downstream fine-tuning procedure.

Table 1 summarizes the prediction heads used during each pretraining stage and downstream application, and implementation details are provided in 'Training procedure, software and hardware' in Methods. These correspond to the task-specific decoder heads shown in Fig. 1. During stage I, only the self-supervised generative head (SS-Gen) is optimized through masked-particle reconstruction, encouraging the backbone to learn task-agnostic event representations. Stage II adds the supervised generation (Sup-Gen) and classification (Cls) objectives to further refine the shared representation. The assignment (Asn) and segmentation (Seg) heads are task-specific and are therefore trained only during downstream fine-tuning.

We isolate the contribution of each stage through the following three model variants:

- EVENET–Scratch (Scratch) is trained from random initialization on each downstream task and serves as the no-pretraining baseline.
- EVENET–SSL (SSL) is initialized from the stage I pretrained checkpoint and therefore isolates the benefit of task-agnostic representation learning before introducing physics-informed supervision.
- EVENET–Full (Full) is initialized from the complete two-stage pretrained model before downstream fine-tuning.

These variants correspond to no pretraining, stage I only and the complete two-stage strategy, respectively. Together, they provide the primary controlled comparison of successive pretraining stages within a common architecture. A complementary classification-only control, described in Supplementary Information, tests conventional supervised transfer using the same backbone and pretraining corpus.

## Generalization across hypothesis grids for discovery

Searches for new physics often require exploring large parameter spaces with many signal hypotheses, each supported by limited simulated data. Models must therefore remain reliable across a wide range of kinematic conditions.

We illustrate this challenge with searches for a heavy new particle decaying into a lighter new particle and a SM Higgs boson ($X\rightarrow YH_{\mathrm{SM}}$), as motivated by several beyond-SM scenarios[31–33]. We evaluate this benchmark using fully simulated CMS Open Data. Because the pretraining samples ('Monte Carlo simulation and Open Data' in Methods) contain no pile-up (multiple interactions within the same bunch crossing), this benchmark tests robustness to domain shifts arising from pile-up and differences in detector response.

We consider cases where the lighter particle decays into two bottom quarks and the Higgs boson decays into a pair of $W$ bosons ($b\bar{b}WW^{*}$), where one $W$ decays to a charged lepton and a neutrino ($\ell\nu$) and the other decays to two quarks ($qq$). Although the pretraining dataset includes SM $H_{\mathrm{SM}}\rightarrow WW^{*}$ events, which constitute approximately 12.8% of the retained pretraining corpus, it contains no $X\rightarrow YH_{\mathrm{SM}}$ topology. Together, this unseen production topology and the distinctive $b\bar{b}WW^{*}$ final state provide a demanding test of whether learned global event representations transfer beyond the pretraining distribution.

We compare EVENET, fine-tuned through its classification head (Fig. 1), with two tabular baselines: XGBoost[34], an established method in high-energy physics, and TabPFN[35,36], a pretrained foundation model for tabular classification. Both operate on fixed-length tables of human-engineered event features, whereas EVENET processes variable-length point clouds of reconstructed objects. For each method, a separate classifier is trained at each point in the ($m_X$, $m_Y$) mass grid. Performance is quantified by the significance-improvement characteristic (SIC)[37],

$$\mathrm{SIC} = \frac{\epsilon_{\mathrm{S}}}{\sqrt{\epsilon_{\mathrm{B}}}}, \qquad (1)$$

where $\epsilon_{\mathrm{S}}$ and $\epsilon_{\mathrm{B}}$ are the signal and background efficiencies. The SIC varies with the chosen discriminant threshold and quantifies the trade-off between signal efficiency and background rejection. To ensure statistical stability, results are reported only for regions with an expected background of at least ten events. Implementation details are provided in 'Generalization across hypothesis grids for discovery' in Methods.

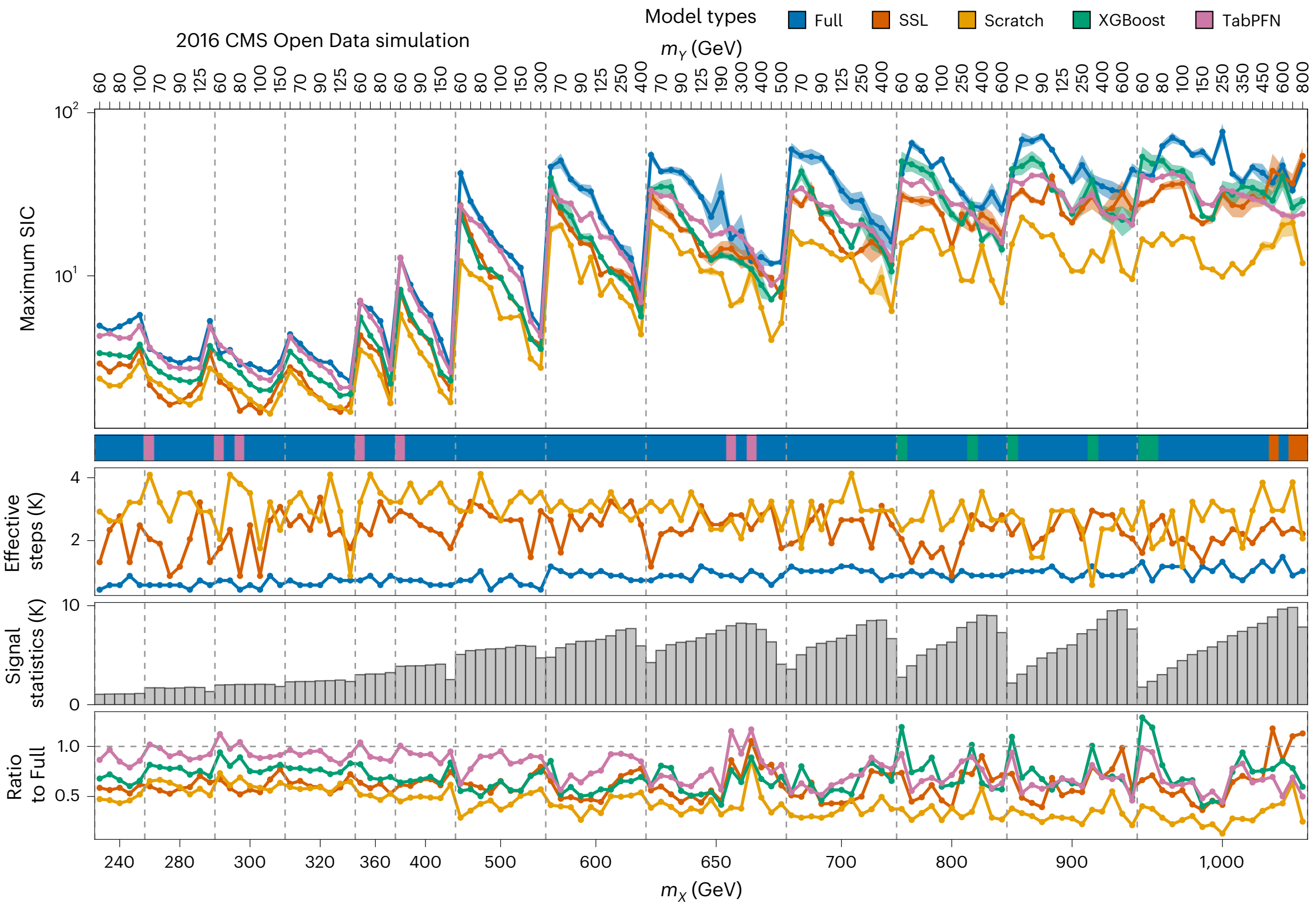

**Fig. 2 | Sensitivity across the ($m_X$, $m_Y$) grid for $X \to YH_{SM}$, $Y \to b\bar{b}$ and $H_{SM} \to WW^*$ using individual training.** The figure contains five panels (from top to bottom). (1) Maximum SIC, evaluated with a minimum background yield of ten events, as a function of $m_X$, with $m_Y$ indicated on the top axis. Markers show the central SIC values and shaded bands indicate 1 s.d. statistical uncertainties obtained by propagating Poisson uncertainties on the signal and background efficiencies into SIC = $\epsilon_S/\sqrt{\epsilon_B}$. (2) Per-bin 'winner' map showing the method that achieves the highest maximum SIC. (3) Effective number of optimization steps required to reach the minimum validation loss. (4) Available signal statistics at each mass point. (5) Ratio of the maximum SIC obtained with each baseline method to that obtained with the pretrained EVENET–Full model.

With individual fine-tuning, EVENET–Full achieves the highest sensitivity across most of the mass grid and remains comparable to the tabular baselines elsewhere (Fig. 2). By contrast, EVENET–Scratch does not consistently surpass these baselines, reflecting the difficulty of training a high-capacity model on low-level inputs with only $O$(1,000–10,000) signal events per mass point. Pretraining also accelerates convergence: EVENET–Full reaches its optimum in approximately 1,000 steps, compared with 3,000 for EVENET–Scratch, reducing downstream optimization effort across the scan. Averaged across the grid, it improves the maximum SIC by approximately 50% relative to EVENET–Scratch, 20% relative to XGBoost and 10% relative to TabPFN. Individual fine-tuning also matches or exceeds all parameterized approaches in Extended Data Fig. 1, indicating an advantage for local optimization in this final state.

## Simulation-to-data transfer with resonance rediscovery

Quarkonium fragmentation within jets provides an important test of quantum chromodynamics (QCD) by probing the transition between perturbative and non-perturbative scales[38–41]. While charmonium production inside jets has been extensively studied[42,43], bottomonium in the anti-isolated regime, where the decay muons are embedded in substantial surrounding hadronic activity, remains largely unexplored. This phase space is difficult to access with standard isolation-based dimuon selections, while relaxing the isolation requirement introduces a large combinatorial background that obscures the $\Upsilon$ signal. These features make anti-isolated $\Upsilon$ production a challenging anomaly-detection benchmark[44]. We therefore search for anti-isolated $\Upsilon \to \mu^+\mu^-$ decays in the 2016 CMS Open Data dimuon dataset[45]. As the analysis is fully data-driven, it provides a direct test of whether the simulation-pretrained EVENET representation transfers to collision data.

Following the CATHODE strategy[44,46], we learn a background model from the dimuon-mass sidebands, interpolate it into the signal region and search for discrepancies with the data. Whereas ref. 44 models low-dimensional engineered observables using a conditional normalizing flow, the diffusion-based EVENET self-supervised generative head models low-level particle-cloud kinematic and vertex information. The generated clouds are calibrated to restore the conditioned dimuon mass and momentum balance, with the required correction providing a diagnostic of generation fidelity.

We then train an XGBoost classifier, following ref. 44, to distinguish the observed signal-region data from the generated background using observables derived from the particle clouds. The classifier is designed to be mass independent to avoid artificially sculpting a resonance. Its output score, $z(\mathbf{x}) = \frac{p_{\rm data}(\mathbf{x})}{p_{\rm data}(\mathbf{x})+p_{\rm bkg}(\mathbf{x})}$, is converted into an estimate of the signal-to-background likelihood ratio

$$\ell(\mathbf{x}) \equiv \frac{z(\mathbf{x}) - (1-\mu)(1-z(\mathbf{x}))}{\mu(1-z(\mathbf{x}))} = \frac{p_{\rm signal}(\mathbf{x})}{p_{\rm bkg}(\mathbf{x})}. \quad (2)$$

Here $p_{\rm data}(\mathbf{x})$, $p_{\rm bkg}(\mathbf{x})$ and $p_{\rm signal}(\mathbf{x})$ denote the probability densities of the classifier inputs $\mathbf{x}$ for the observed signal-region data, the estimated background and the signal component, respectively, and $\mu$ is the signal fraction estimated from the unweighted dimuon-mass spectrum. This quantity serves as the most powerful mass-independent test statistic[47,48]. Events are weighted by $\ell(\mathbf{x})$, and the local significance is extracted from the reweighted dimuon-mass spectrum. Each pseudo-experiment resamples the generated background and retrains the classifier with independent randomization; performance is quantified by the median significance and its spread across pseudo-experiments. Further details are provided in 'Simulation-to-data transfer with resonance rediscovery' in Methods.

We analyse both opposite-sign (OS) and same-sign (SS) dimuon channels: a physical $\Upsilon \to \mu^+\mu^-$ resonance appears only in the OS channel, while the SS channel provides a background-only control. Configurations are denoted train → test, indicating the channels used for training and evaluation, respectively. In the primary OS → OS configuration, the generator is trained on 10,640 OS sideband events, and the classifier is trained and cross-validated on 1,692 OS signal-region events.

In the OS → OS configuration, calibrated EVENET–Full reaches a median local significance of $7.6^{+0.4}_{-0.4}\sigma$, exceeding the published $6.4\sigma$ (ref. 44; Fig. 3a). EVENET–SSL achieves a comparable $7.4^{+0.4}_{-0.4}\,\sigma$, whereas EVENET–Scratch reaches only $1.5^{+0.2}_{-0.2}\sigma$ after calibration. The median calibration corrections are 11.6% for EVENET–Full and 13.3% for EVENET–SSL, compared with 74.4% for EVENET–Scratch (Fig. 3b). Together with their higher post-calibration significances, these smaller corrections indicate that the pretrained representations are more robust under physical consistency constraints.

The representative OS → OS spectrum exhibits a clear excess in the $\ell$-reweighted signal region (Fig. 3c). By contrast, the OS → SS and SS → SS control configurations are compatible with zero significance (Fig. 3a,d), providing no evidence of an artificially sculpted resonance. The SS → OS configuration gives a small non-zero significance because physical dimuon resonances are present in the OS evaluation data.

Taken together, these results show that although anomaly detection is feasible in this low-statistics regime[44], learning a high-fidelity low-level particle-cloud distribution remains challenging from limited sideband data. The contrast between the pretrained and scratch models therefore highlights the importance of pretraining in mitigating this challenge.

### Physics-structured multi-task synergy and robustness

To test whether a single pretrained event representation can support structurally different collider analyses, we consider two multi-head downstream tasks: an out-of-distribution exotic Higgs search and a high-statistics precision quantum-correlation measurement in dileptonic top-quark pair events. In both cases, performance depends on jointly optimizing heads that address complementary aspects of event reconstruction, allowing us to test whether pretraining supports multiple structured objectives. Because detector-response mismodelling can bias inference, we also assess robustness by applying representative detector-level variations at evaluation time without retraining, isolating representation stability under detector-induced distribution shifts from re-optimization effects. In this section, training sizes are reported relative to task-specific reference datasets: 200,000 events for the exotic Higgs search and 1 million events for the quantum-correlation measurement. These reference sizes reflect typical LHC analysis statistics. All methods use validation-based early stopping and comparisons use each model's best-performing checkpoint.

**Exotic Higgs search.** Direct searches for exotic Higgs decays remain a central probe of physics beyond the SM, with strong motivation for rare non-standard final states[49,50]. We study $H_{\rm SM} \to aa \to 4b$ (refs. 51–53), where $a$ is a new pseudoscalar particle with mass $m_a$, and sensitivity is limited by large multijet backgrounds from QCD, combinatorial jet misassignment and flavour mistagging.

This channel is absent from pretraining and therefore provides an out-of-distribution transfer test. Motivated by previous observations that auxiliary heads can improve classifier sensitivity[54], we fine-tune EVENET using its classification, particle-assignment and particle-segmentation heads (Fig. 1c) in three configurations: Cls, Cls + Asn and Cls + Seg. We report the maximum SIC, corresponding to the classifier threshold that maximizes it. The downstream task and training set-up are described in 'Exotic Higgs search' in Methods.

As shown in Fig. 4a, EVENET–Full consistently outperforms both EVENET–Scratch and SPANet[54,55], a state-of-the-art transformer-based model for joint particle assignment and classification in high-energy physics, across reference-dataset fractions. At the representative $m_a$ = 30 GeV point, EVENET–Full (Cls) reaches SIC = 4.1, versus 1.6 for scratch and 1.4 for SPANet. With 5% of the reference dataset, it reaches SIC = 1.7, exceeding both baselines' peak values at 150% of the reference dataset. This advantage persists across intermediate and large fractions.

Robustness under detector-response variations is evaluated with 100 pseudo-experiments under the Cls + Asn configuration by varying the jet transverse momentum with a fractional jet-energy-scale (JES) shift, $\alpha_{\rm JES}$. Figure 4c shows that EVENET–Full remains substantially more stable than scratch: the 68% and 95% central intervals of the normalized SIC shift, defined as $({\rm SIC} - \langle {\rm SIC}\rangle)/\sigma_{\rm SIC}$, where $\langle {\rm SIC}\rangle$ and $\sigma_{\rm SIC}$ denote the mean SIC across variations and the propagated statistical uncertainty for each variation, respectively, are 0.21 and 0.34 for EVENET–Full, compared with 0.71 and 1.42 for scratch. These results indicate that pretraining regularizes the shared event representation, improving both sensitivity and robustness to correlated detector-induced perturbations in a multi-head setting.

**Top-quark pair quantum-correlation measurement.** Quantum correlations in top-quark pair production provide a precision benchmark limited by reconstruction accuracy rather than data availability. In dileptonic final states, both top quarks decay to a charged lepton and an undetected neutrino, while angular correlations of the leptons encode the pair's spin correlations[56,57]. These correlations define entanglement-sensitive observables and have been recently measured by ATLAS and CMS[58,59].

Full event reconstruction remains the dominant limitation because the two neutrinos and finite detector resolution leave the kinematic system under-constrained. Measurement precision therefore depends critically on neutrino momentum inference and object assignment. We address this using physics-structured particle-assignment and supervised generative heads that jointly resolve combinatorial ambiguities and infer the missing neutrino momenta.

Spin-correlation coefficients $C_{ij}$ are reconstructed using the decay-angle formalism described in Methods (equations (8) and (9)), and combined into the entanglement-sensitive observable $\Delta = |C_{kk} + C_{rr}| - C_{nn} - 1$, for which $\Delta > 0$ indicates quantum entanglement[56]. Measurement performance is quantified using the precision metric $\sigma_\Delta/\Delta$. All evaluated configurations yield $\Delta > 0$, ensuring that the precision is well defined. Details of the dataset and evaluation procedure are given in 'Top-quark pair quantum-correlation measurement' in Methods.

With ~1.5% of the reference dataset, EVENET–Full reaches a measurement precision of 3.88%, compared with 6.20% for SSL only and 5.58% for scratch training. Precision improves to 3.37% at ~15% and saturates near 3.33%, approximately 37% better than the 5.26% benchmark in ref. 60. At the largest reference fractions, the separation narrows to 3.33% for EVENET–Full versus 3.44–3.55% for scratch and SSL. This narrowing reflects a high-statistics regime where further gains are no longer statistics limited.

Because precision top-quark pair measurements are typically systematics dominated, we further test detector-level stability with JES and $E_T^{\rm miss}$ variations applied at evaluation time, where $E_T^{\rm miss}$ is the magnitude of the negative vector sum of the transverse momenta of all reconstructed objects. As shown in Fig. 5c,d, EVENET–Full keeps the spread in the precision shift, defined as $\sigma_\Delta/\Delta - \langle\sigma_\Delta/\Delta\rangle$, within $5\times10^{-4}$ at 68% confidence and $1\times10^{-3}$ at 95% confidence for both sources, while scratch fluctuations are about 5 times larger. This robustness gap under key reconstruction systematics indicates that pretraining improves both central performance and stability of precision extraction.

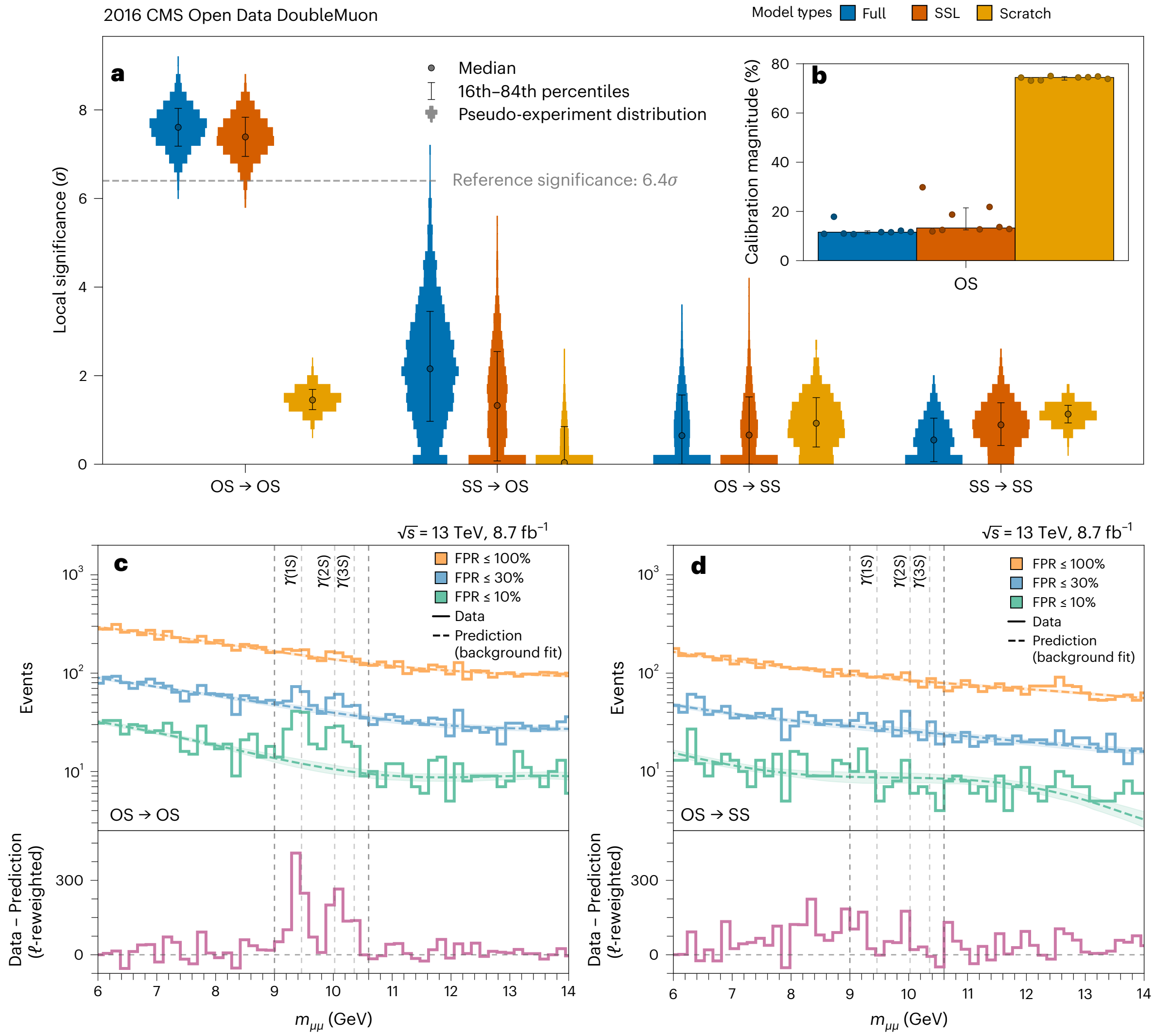

**Fig. 3 | Collision-data anomaly detection using dimuon particle-cloud generation. a**, Median ℓ-reweighted local significance for four train-to-test configurations: opposite-sign to opposite-sign (OS → OS), same-sign to opposite-sign (SS → OS), opposite-sign to same-sign (OS → SS) and same-sign to same-sign (SS → SS). Blue markers denote EVENET–Full, yellow markers denote EVENET–Scratch and orange markers denote EVENET–SSL. The horizontal dashed line marks the published reference significance. Pseudo-experiments are pooled across 8 independently trained generator ensembles, with 200 resamplings per ensemble. Each model has $n = 1{,}600$ pseudo-experiments per configuration. Data are presented as the median ℓ-reweighted local significance with error bars showing the 16th–84th percentile interval of the pooled distribution. Mirrored histograms show the pooled distributions in common bins of width $0.2\sigma$; widths are proportional to bin counts and normalized separately to the same maximum width. Each pseudo-experiment resamples 20,000 generated background events and retrains the classifier. **b**, Calibration magnitude required to restore exact dimuon two-body kinematics in the opposite-sign region. Each of $n = 8$ independently trained generator ensembles contributes its mean relative four-momentum adjustment; bars show the median of these 8 values, with overlaid points showing the individual ensemble values and error bars spanning their 16th–84th percentiles. **c**,**d**, Representative dimuon-mass spectra obtained with EVENET–Full at fixed false-positive-rate (FPR) operating points (top), together with the corresponding ℓ-reweighted distributions (bottom), for the signal-sensitive OS → OS (**c**) and control OS → SS (**d**) configurations. Sideband fits are interpolated into the signal region. Shaded bands around the fitted backgrounds show the 16th–84th percentile intervals of predictions obtained by sampling the fitted parameter covariance. Coloured dashed curves show the fitted background predictions, while vertical grey dashed lines indicate the signal-region boundaries and the nominal masses of the ϒ(1S), ϒ(2S) and ϒ(3S) resonances. The displayed pseudo-experiment in each panel is the one with an ℓ-reweighted significance closest to the median of the corresponding train-to-test configuration.

## Training convergence

An essential property of reusable foundation models is efficient task-specific adaptation after pretraining. We characterize downstream convergence using the minimum validation loss and the number of effective training steps required to reach it. Effective steps equal the number of training examples processed before the minimum loss, divided by the global batch size, and therefore correspond to optimizer updates.

On the quantum-correlation task, EVENET–Full reaches a validation loss of 1.11 within $\mathcal{O}(200)$ effective steps, compared with 1.46 for SSL only and 1.87 for scratch training in the low-statistics regime (Fig. 5b). At larger sample sizes, it reaches the optimal plateau of 0.86 with nearly an order of magnitude fewer updates than scratch.

The exotic Higgs benchmark shows a similar pattern. In the representative Cls + Asn configuration, EVENET–Full reaches losses of 0.40–0.45 within $\mathcal{O}(10^2)$–$\mathcal{O}(10^3)$ effective steps, whereas SSL only converges more slowly and scratch requires $\mathcal{O}(10^3)$–$\mathcal{O}(10^4)$ steps (Fig. 4b).

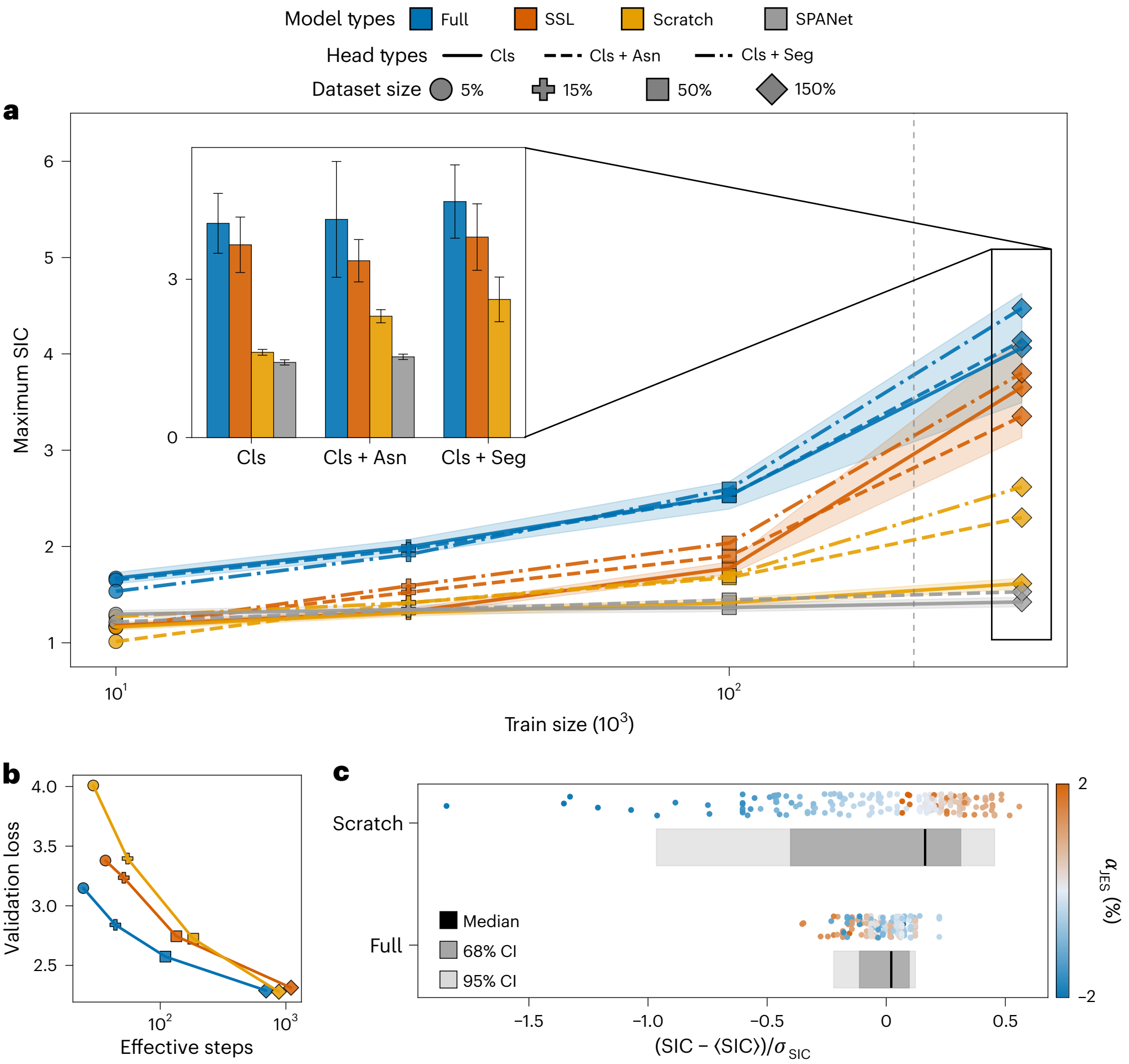

**Fig. 4 | Out-of-distribution generalization to exotic Higgs boson decays.** **a**, Maximum SIC as a function of training sample size, expressed relative to a reference dataset containing 200,000 events. Markers show the central SIC values; error bars and shaded bands indicate 1 s.d. statistical uncertainties obtained by propagating Poisson uncertainties on the signal and background efficiencies into SIC $= \epsilon_S/\sqrt{\epsilon_B}$. SPANet is omitted from the Cls + Seg comparison because it does not support segmentation tasks. The inset compares the maximum SIC across models and task-head configurations at 150% of the reference dataset. **b**, Validation loss as a function of the effective number of training steps for the Cls + Asn configuration, illustrating representative training dynamics. **c**, Normalized shift in the maximum SIC, $(\mathrm{SIC} - \langle \mathrm{SIC} \rangle)/\sigma_{\mathrm{SIC}}$, evaluated under JES variations sampled from a Gaussian prior with a width of 1% for the Cls + Asn configuration ($n = 100$ systematic variations per model). Each point in **c** corresponds to an individual systematic variation; the vertical line and shaded bands indicate the median response and the central 68% and 95% intervals, respectively.

Across both benchmarks, EVENET–Full reaches equal or lower validation loss with fewer downstream optimizer updates. This comparison excludes the one-time pretraining cost reported 'Training procedure, software and hardware' in Methods.

## Discussion

In this study, we introduce EVENET, a foundation model for high-energy physics analysis. Across four downstream settings spanning precision measurement, new-physics searches and collision-data anomaly detection, EVENET–Full attains performance comparable to or better than task-specific state-of-the-art baselines and the pretrained tabular foundation model TabPFN. Within the common architecture, this two-stage pretraining model also matches or outperforms the variants with no pretraining (EVENET–Scratch) or one-stage pretraining (EVENET–SSL), often requiring substantially fewer optimization steps during fine-tuning (Fig. 2 and 'Training convergence'). Unlike previous particle-based pretraining approaches focused on discriminative tasks[28,29], EVENET uses hybrid discriminative-generative pretraining to learn a shared particle-cloud representation supporting both inference-driven analyses and generative reconstruction across diverse physics processes and data domains. This hybrid pretraining strategy also highlights a central lesson of this work: whereas SSL alone is often sufficient in other domains, pretraining on collider data with explicit physical structure can exploit unambiguous simulation truth labels as strong inductive signals. In our experiments, incorporating this supervision improves upon SSL-only pretraining.

Across disparate downstream regimes, a shared pattern emerges: a single pretrained backbone can be adapted efficiently across heterogeneous inference settings without loss of physics performance. In searches for heavy Higgs bosons, EVENET uses a shared pretrained initialization across scan points, enabling efficient per-hypothesis fine-tuning and avoiding the need to train models from scratch at each mass hypothesis, while matching or exceeding state-of-the-art and tabular foundation-model baselines across strongly varying kinematics. This advantage is reflected in the training dynamics: pretrained variants reach lower validation loss with substantially fewer optimization steps during downstream adaptation. The substantial one-time pretraining cost can be amortized through reuse of the backbone across multiple analyses.

A complementary effect appears when downstream objectives depend explicitly on event topology or multi-step inference. In exotic Higgs decays, models trained from scratch benefit from auxiliary assignment or segmentation heads, reflecting the discriminative value of resonance structure. For EVENET−Full, however, this auxiliary gain is reduced: high sensitivity is already achieved using the classification head alone. These observations suggest that, with pretraining, human-designed auxiliary heads become less essential, lowering reliance on task-specific architectural augmentation during fine-tuning. Moreover, stable end-to-end training is maintained even when the downstream loss couples multiple inference targets. We observe this robustness not only in the new-physics search setting above but also in quantum-correlation measurements, which require the joint inference of resonance assignment and neutrino reconstruction, and in which measurement uncertainties are often driven by systematic effects rather than statistical fluctuations. Taken together, these aspects highlight the potential of the pretrained backbone to tackle complex, analysis-grade problems.

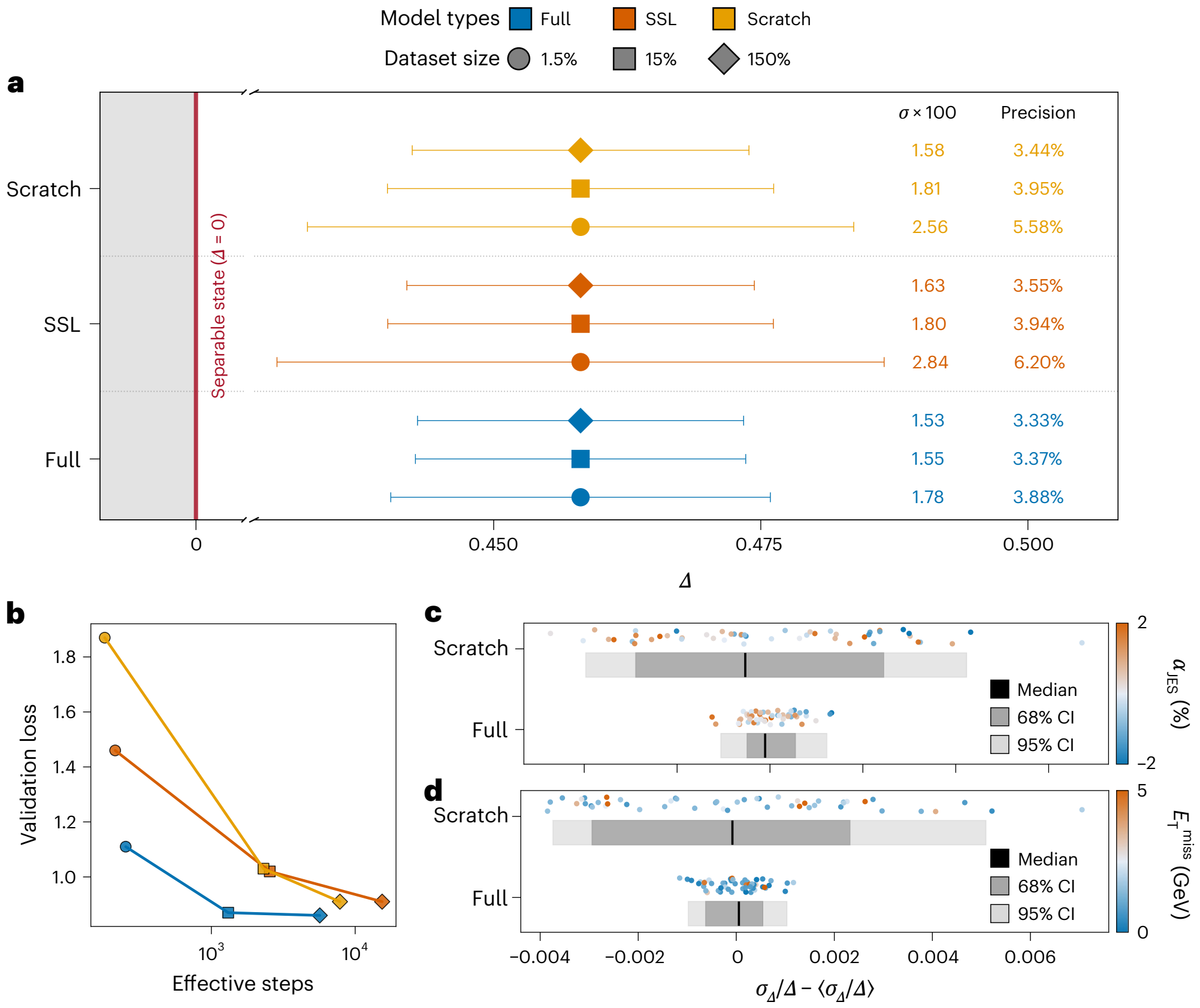

**Fig. 5 | Precision measurement of quantum correlations in top-quark pairs.** **a**, Point estimates of the entanglement-sensitive observable $\Delta$, shown for training sample sizes expressed relative to a reference size of $10^6$ training events, representative of top-quark analyses. Evaluation uses $n = 445{,}673$ selected events from a test sample of 8,195,408 events (Extended Data Table 1). Central values are the extracted $\Delta$ measurements; error bars show $\pm\sigma_\Delta$, where $\sigma_\Delta$ is the 1 s.d. statistical uncertainty propagated from Poisson counting uncertainties in the decay-angle product distributions used to reconstruct the spin correlations. The columns alongside the estimates report $100\sigma_\Delta$ and the relative precision $\sigma_\Delta/\Delta$ expressed as a percentage. Circles, squares and diamonds denote training fractions of 1.5%, 15% and 150%, respectively. The red line marks the separability boundary $\Delta = 0$; the horizontal axis is broken between this boundary and the estimates. **b**, Validation loss as a function of the effective number of training steps, illustrating representative training dynamics. **c**, Shift in the measured relative precision of $\Delta$ under JES variations sampled from a Gaussian prior with a width of 1% ($n = 50$ systematic variations per model). **d**, Shift in the measured relative precision of $\Delta$ under fluctuations in the soft component of the missing transverse momentum, generated from a log-normal distribution and truncated to $0 \leq \Delta E_T^{\text{miss}} \leq 5$ GeV ($n = 50$ systematic variations per model). In **c** and **d**, the horizontal axis shows the mean-subtracted relative precision, $\sigma_\Delta/\Delta - \langle\sigma_\Delta/\Delta\rangle$. Each point represents one systematic variation; colour indicates the applied JES variation in **c** and the soft missing-transverse-momentum variation magnitude in **d**. Vertical lines and dark and light shaded bands indicate the median response and the central 68% and 95% intervals, respectively.

This cross-task coherence extends to self-supervised generative deployment on collision data. In anomaly detection, EVENET−Full maintains stable performance across detector and kinematic regimes, including low-statistics settings, surpassing the published benchmark after calibrated reconstruction (median significance $7.6^{+0.4}_{-0.4}\sigma$ versus $6.4\sigma$ in ref. 44). Models trained from scratch instead show a lower post-calibration significance and require substantially larger kinematic corrections.

Together, these results support the use of a unified event-level foundation model as a practical substrate for both discriminative and generative analyses, consolidating traditionally multi-stage pipelines into a single backbone while retaining robustness to the systematic variations evaluated here.

Despite these advances, several challenges remain. Although EVENET incorporates noise-tolerance training, it is not yet a fully uncertainty-aware framework: explicit modelling and propagation of detector systematics and theoretical uncertainties remain necessary for rigorous statistical interpretation. In addition, while multi-head architectures are effective in practice, predictions from different heads are not explicitly constrained to be mutually consistent, leaving scope for future approaches that jointly enforce coherence across classification, assignment and generative reconstruction outputs. Transfer to collision data is encouraging, but systematic stress tests under substantial simulation-data mismodelling have not yet been performed. Finally, although the shared particle-cloud representation captures event structure and kinematics, a detailed analysis of latent-space organization and interpretability is deferred to future work.

In conclusion, EVENET illustrates the promise of event-level foundation models as a flexible framework for a broad class of analyses. Deployment in LHC analyses will provide a decisive test of its ability to streamline end-to-end inference, particularly when integrated with object-level frameworks such as OmniLearned[27]. Looking ahead, coupling EVENET to differentiable detectors[14] or fast-simulation frameworks[61] could enable more tightly integrated analysis pipelines, supporting gradient-based optimization from reconstructed observables towards theory parameters. At the same time, the learned shared particle-cloud representation suggests opportunities for more interpretable embeddings and anomaly-sensitive representations. Taken together, these results position EVENET as a concrete step towards foundation models for collider physics, with a clear near-term validation milestone and a restrained longer-term outlook.

# Methods

In this section, we give implementation-level details on dataset preparation, model architecture and training set-up. We also clarify how the downstream analyses were carried out and formalize the performance metrics used in our studies.

## Monte Carlo simulation and Open Data

To generate datasets for both pretraining and downstream studies, we used large-scale Monte Carlo simulations covering a broad spectrum of SM processes. SM hard-scattering processes are generated with MADGRAPH5_aMC@NLO 3.5.5[62], covering multijet backgrounds, top-quark final states and electroweak boson final states across a range of jet multiplicities. Parton showering and hadronization are modelled with PYTHIA 8[63], and detector responses are approximated using Delphes[64] with a generic detector configuration. All samples produced with this Delphes chain correspond to proton–proton collisions at a centre-of-mass energy of $\sqrt{s} = 13$ TeV and are generated without pile-up.

Generator-level preselections define the phase-space support of the event representation and suppress low-energy or out-of-acceptance configurations. Reconstructed jets and charged leptons are required to satisfy a transverse momentum $p_T > 10$ GeV, with charged leptons restricted by imposing a pseudorapidity requirement of $|\eta| < 2.5$, where $\eta$ represents the angle of a particle relative to the beam pipe. Additional process-dependent filters enforce minimal object multiplicities consistent with the underlying hard process; for example, QCD samples require at least two jets with $p_T > 20$ GeV, while top-quark and electroweak samples impose lepton, jet and $b$-jet multiplicity requirements matching the target final states.

The pretraining corpus comprises approximately $3 \times 10^9$ simulated SM events, from which about $5 \times 10^8$ events are retained after preselections. These samples span the dominant LHC background and signal-like processes and are used exclusively to initialize the shared event representation before task-specific optimization.

Four downstream datasets are used and are divided by data provenance. CMS Open Data are used for the hypothesis-grid study ($X \to YH_{SM}$ samples) and for anomaly detection using the 2016 DoubleMuon dataset[45]. The remaining studies use Delphes-simulated samples: dileptonic top-quark pair events for quantum-correlation measurements, and $H_{SM} \to aa \to 4b$ signal samples with $m_a = 30$ GeV together with multijet backgrounds containing two or four $b$ jets. All datasets are summarized in Extended Data Table 1.

Data-splitting strategies are defined separately for each stage and downstream task to reflect their intended use and to avoid information leakage. For pretraining, simulated events are randomly split into training (70%) and validation (30%) subsets. No dedicated test set is used at this stage, as model evaluation is performed exclusively on downstream tasks. For the hypothesis-grid study, datasets are partitioned into two disjoint subsets: one used for training and validation and the other reserved for final evaluation. For anomaly detection, fivefold cross-validation is used to maximize statistical usage while maintaining strict separation between training and evaluation data in each fold. For the quantum-correlation and exotic Higgs studies, independent datasets are generated for training and testing, ensuring statistical orthogonality and eliminating overlap between training and evaluation samples.

## Model architecture

The transformer architecture provides a permutation-equivariant mechanism for modelling sets of physics objects through attention-based interactions[65]. To incorporate local geometric structure, we augment the transformer with $k$-nearest-neighbour message passing, yielding a hybrid point–edge transformer architecture[26]. The resulting point–edge transformer encoder serves as a shared backbone that maps reconstructed objects into a unified event-level representation, which is consumed by multiple task-specific heads (Fig. 1).

Reconstructed physics objects, including jets, electrons and muons, together with their kinematic properties, flavour-tagging information, lepton identification and charge, are represented as particle clouds and provided as input to the backbone. During pretraining, a 10% feature dropout is applied by randomly masking input features on a per-sample basis to discourage over-reliance on specific features and to improve generalization to downstream tasks. The backbone supports both discriminative and generative objectives within a single representation space. Generative objectives are formulated using diffusion models[66], which reconstruct clean inputs from progressively noised versions. The diffusion-time parameter $t$ controls the noise level and defines a continuum between clean inputs ($t = 0$) and increasingly perturbed representations ($t > 0$). Discriminative and generative objectives are evaluated on representations sampled along this noise trajectory, enabling a consistent interface for both regimes within the shared backbone.

The classification head predicts the physics process category of each event. During pretraining, the classification head predicts one of the ten physics-process categories listed in Extended Data Table 1: QCD, $t\bar{t}$, $t\bar{t}W$, $t\bar{t}Z$, $W$ + jets, $Z$ + jets, $WW$, $ZZ$, $WZ$ and $H_{SM}(WW^*)$. Variations in decay channels within each process (for example, hadronic or leptonic decays) share the same process label. Classification is performed on representations sampled across diffusion-time steps, including both clean and noise-perturbed inputs. For downstream classification tasks, compatible layers of the pretrained classification head are retained, while the final class-dependent layers are replaced and randomly initialized to match the required output dimensionality. Unlike during pretraining, downstream classification is performed only on clean inputs.

The classification loss is defined as $\mathcal{L}_{\text{class}} = \alpha^2(t)\ \text{CE}\,(y, \hat{y}_{\text{pred}})$, where $\hat{x}_{\text{perturbed}} = \alpha(t)x + \sigma(t)\epsilon$, with $\lim_{t\to 0}\alpha(t) = 1$ and $\lim_{t\to 1}\alpha(t) = 0$, $\epsilon$ denoting Gaussian noise, $\hat{y}_{\text{pred}}$ the prediction obtained from the noise-perturbed input and CE the cross-entropy. The factor $\alpha^2(t)$ weights contributions at different perturbation levels.

The assignment head predicts object-to-decay associations within a hierarchical representation consisting of resonances and their decay products. For each resonance $r$, an assignment tensor encodes the probabilities of mapping the $n$ reconstructed particle-cloud objects to the $d_r$ decay products and has dimension $n^{d_r}$. Multiple transformer-based heads incorporating the symmetry tensor attention framework[54,55] predict resonance reconstructability and assignment tensors while respecting permutation symmetries among identical decay products. Resonances sharing identical decay structures are processed by a shared head. Resonance attributes, such as masses, are provided as conditional inputs so the model can respect and exploit symmetries between resonances.

The assignment loss follows the SPANet formulation[54,55], combining a permutation-invariant cross-entropy loss for object assignments with a cross-entropy loss for resonance-reconstructability prediction. Non-reconstructable resonances are masked from the assignment term. The loss is reweighted to account for imbalances among reconstructability configurations.

The segmentation head formulates resonance prediction as an object-detection problem following the DETR framework[67] and its mask-based extensions[68]. A set of learnable queries is processed by a transformer decoder with cross-attention to the encoded particle-cloud representation and refinement via self-attention. Each query corresponds to a resonance candidate and is mapped to a class label and a segmentation mask over reconstructed objects.

The segmentation loss follows the DETR-style set-prediction formulation[67,68]. During training, predictions are first associated with targets using Hungarian matching[69], ensuring permutation invariance among resonance candidates. The matching cost includes classification and mask components. After matching, the loss combines a class-reweighted classification term with focal[70] and dice[71] mask terms, applied only to matched resonance targets.

The unsupervised generative head defines a self-supervised pretraining objective based on diffusion-conditioned reconstruction. A masking module perturbs a subset of particle-cloud elements and provides a diffusion-time embedding that encodes the noise level for each perturbed point. The generative head conditions on the unperturbed portion of the event representation to reconstruct the perturbed inputs, corresponding to an inverse diffusion denoising task. Fully perturbing the particle cloud recovers the unconditional generation setting within the same formulation.

The supervised generative head follows the same diffusion-based formulation as the self-supervised reconstruction objective but targets invisible event components. Inspired by the use of conditional diffusion models for neutrino momentum reconstruction in ref. 72, noised tokens representing invisible particles are injected into the particle cloud and reconstructed conditioned on the observed objects. The missing transverse momentum is included as a global event-level conditioning variable, allowing the model to learn the relationship between the event-wide transverse momentum imbalance and the invisible components. From the perspective of the shared backbone, supervised reconstruction of invisible components and self-supervised masked reconstruction correspond to the same diffusion-conditioned denoising operation, differing only in the choice of perturbed tokens. Task-specific decoders define the reconstruction targets in each case. Both objectives use the loss $\mathcal{L}_{\mathrm{generation}} = \| \mathbf{v} - \mathbf{v}_{\mathrm{pred}} \|_2^2$, where $\mathbf{v}$ and $\mathbf{v}_{\mathrm{pred}}$ denote the target and predicted diffusion vectors, respectively.

## Training procedure, software and hardware

Model training was performed on the Perlmutter supercomputer at National Energy Research Scientific Computing Centre (NERSC) using 512 NVIDIA A100 graphics processing units (GPUs) and 16,384 central processing unit cores. The implementation is based on PyTorch 2.6[73], with PyTorch Lightning for training control and the Ray[74] distributed runtime for multi-node orchestration.

Pretraining followed a two-stage curriculum. In the first stage, the network was trained in a fully self-supervised configuration with masked inputs. Only the self-supervised generative head was active, and the masking probability was increased progressively from 30% to 100% during training. In the second stage, all weights from the first stage were reused, and the network was trained jointly with the classification, supervised generative and self-supervised generative heads active. This configuration defines the EVENET pretraining set-up used for all downstream analyses and is summarized in Table 1. The model-training component of stage I and stage II pretraining required approximately $\mathcal{O}(10^5)$ GPU-hours on 512 NVIDIA A100 GPUs, including both stage I and stage II pretraining. This estimate refers to model training only and excludes event generation and preprocessing. The pretraining corpus comprises approximately 4 TB of raw simulation, which is reduced to approximately 180 GB after preprocessing into the parquet format used for training. Downstream fine-tuning tasks typically required between a few and tens of GPU-hours depending on dataset size and task complexity. Although the upfront pretraining cost is substantial, it is incurred only once, and the resulting pretrained model can be reused across multiple downstream analyses.

Losses from all active heads were summed directly, without gradient surgery or adaptive reweighting. Each loss term was normalized by the number of contributing processes or particles to maintain comparable magnitudes across heads. Training used the LION optimizer[75] with exponential moving average updates and a cosine warm-up and decay schedule. Base learning rates were $1.8 \times 10^{-5}$ and $1.2 \times 10^{-5}$ for the first and second stages, respectively. The effective learning rate scaled with $\sqrt{N_{\mathrm{GPU}} = 512}$, and the weight decay followed inverse scaling, $10^{-3}/\sqrt{512}$.

During downstream fine-tuning, encoder and decoder parameters were optimized jointly. The encoder learning rate was set to one order of magnitude lower than the decoder learning rate.

## Downstream applications

**Generalization across hypothesis grids for discovery.** Simulated events are taken from the CMS Open Data release corresponding to proton–proton collisions at $\sqrt{s} = 13$ TeV recorded in 2016[76–80]. Only fully simulated samples using the CMS detector are considered, and the study is restricted to the resolved regime with $m_X \le 1$ TeV. Signal events correspond to the process $X \to YH_{\mathrm{SM}} \to b\bar{b}WW^{*} \to b\bar{b} + \ell\nu qq$, generated at 121 mass points spanning the $(m_X, m_Y)$ plane. The complete list of CMS Open Data signal datasets and the corresponding DOI links are provided in the supplementary signal-sample list (Supplementary Tables 1–4). Included SM backgrounds comprise semileptonic top-quark pair production, Drell–Yan production with heavy-flavour jets, and single-top processes in the $s$, $t$ and $tW$ channels. The simulated samples used in this study and their production cross-sections are summarized in Extended Data Table 2.

Events are reconstructed in the resolved topology and are required to contain exactly one charged lepton and no hadronically decaying $\tau$ candidates. Electrons must satisfy $p_{\mathrm{T}} > 35$ GeV and $|\eta| < 2.5$, while muons must satisfy $p_{\mathrm{T}} > 30$ GeV and $|\eta| < 2.4$. Jets are reconstructed with $p_{\mathrm{T}} > 20$ GeV and $|\eta| < 2.4$, and hadronically decaying $\tau$ candidates are required to satisfy $p_{\mathrm{T}} > 30$ GeV and $|\eta| < 2.3$. To avoid double counting, $\tau$ candidates within $\Delta R < 0.4$ of selected electrons or muons are removed, and jets within $\Delta R < 0.4$ of selected electrons, muons or remaining $\tau$ candidates are removed. Selected events are required to contain at least two $b$-tagged jets and at least two light jets, and to satisfy $\Delta R(\ell, j_{\mathrm{near}}) < 1.6$, where $j_{\mathrm{near}}$ denotes the light jet closest to the lepton. A topness discriminator based on a kinematic reconstruction of the semileptonic top-quark hypothesis is applied, defined as the minimum $\chi^2$ over all jet assignments to the leptonic and hadronic top candidates, with events required to satisfy topness >3.

The hypothesis-grid study is formulated as a two-class signal-versus-background classification problem, with all background processes combined into a single background class. During both training and evaluation, these background processes are normalized to their expected yields using their production cross-sections. For comparison, XGBoost[34] and TabPFN[35,36] are used. Both baselines are trained on the same engineered feature set constructed from reconstructed objects and composite candidates, including kinematic variables, invariant and transverse masses of composite systems, angular correlations, and topness-related variables. For each event, two alternative hadronic-$W$ hypotheses (on-shell and off-shell) are considered, and variables associated with both hypotheses are included in the tabular input, as summarized in Extended Data Table 3. XGBoost follows the standard gradient-boosted decision tree training procedure. TabPFN uses version v2.5[36]. The number of training events for TabPFN is capped at 50,000 owing to model constraints; when downsampling is necessary, we downsample only background events to preserve signal statistics. The default ensemble of eight models with class balancing is used.

EVENET is evaluated under two training configurations: (1) individual training at each mass point and (2) parameterized training using

$(m_X, m_Y)$ as additional input features. Its input variables follow those used during pretraining and are summarized in Extended Data Table 3. All models are trained and evaluated on independent $X \to YH_{SM}$ samples with 1,000–10,000 signal events per mass point, depending on the mass hypothesis. For individual mass-point training, most backbone parameters are frozen, with only the final layers partially unfrozen and updated at a learning rate set to 0.3 times the base learning rate. Following ref. 81, lightweight adapters are inserted between transformer blocks. This parameter-efficient set-up reduces the number of trainable parameters and the downstream fine-tuning cost for each mass point while retaining task-specific adaptation capacity.

Performance is evaluated using the signal SIC[37]. Efficiencies are computed from threshold scans over the classifier score. When reporting the maximum SIC, a minimum expected background yield of ten events is required.

**Simulation-to-data transfer with resonance rediscovery.** The analysis uses the 2016 CMS Open Data `DoubleMuon` dataset recorded at $\sqrt{s} = 13$ TeV. Events are selected using the `HLT_TrkMu15_DoubleTrk-Mu5NoFiltersNoVtx` trigger. Offline selections require $p_T(\mu_1) > 17$ GeV and $p_T(\mu_2) > 8$ GeV, with both muons satisfying the tight muon identification criteria and the anti-isolation requirement $\mathrm{Iso}_{0.4} > 0.55$. Here $\mathrm{Iso}_{0.4}$ denotes the relative isolation observable evaluated within $\Delta R = 0.4$ of the muon, where $\Delta R$ is the angular distance in pseudorapidity–azimuth space. Events are divided into OS and SS dimuon channels according to the muon charges. The mass range is divided into a left sideband ($5 < m_{\mu\mu} < 9$ GeV), a signal region ($9 < m_{\mu\mu} < 10.6$ GeV) and a right sideband ($10.6 < m_{\mu\mu} < 16$ GeV).

The background density is factorized as

$$p_{\mathrm{bkg}}(\mathbf{x}, m) = p_{\mathrm{bkg}}(\mathbf{x}|m)\, p_{\mathrm{bkg}}(m), \tag{3}$$

where $m = m_{\mu\mu}$ and $\mathbf{x}$ denotes the remaining event observables. The mass density $p_{\mathrm{bkg}}(m)$ is obtained from a quintic-polynomial fit to the sidebands using a relative bin width of 1.5%, consistent with the dimuon-mass resolution inferred from $J/\psi$ fits[44]. The conditional distribution $p_{\mathrm{bkg}}(\mathbf{x}|m)$ is learned using EVENET.

An ensemble of five diffusion-based EVENET generators is trained using a fivefold partition of the sideband data, with the generated samples from all folds combined to form the background estimate. The EVENET–Full, EVENET–SSL and EVENET–Scratch configurations are evaluated using the same downstream analysis procedure. For the pretrained variants, the backbone learning rate is set to 10% of the base learning rate during fine-tuning, whereas EVENET–Scratch uses a uniform learning rate.

Generation proceeds in two stages. First, the event-level observables $H_T$, the scalar sum of the transverse momenta of all jets in the event and $\Delta R(\mu_1, \mu_2)$ are generated conditionally on the dimuon mass. The full dimuon particle cloud is then generated conditionally on the dimuon mass and these intermediate observables, with each muon represented by its transverse momentum $p_T$, energy $E$, pseudorapidity $\eta$, azimuthal angle $\phi$ and three-dimensional impact parameter (IP3D). This hierarchical formulation is intended to reduce the complexity of directly generating the full particle cloud by first modelling the global event structure and then conditioning the detailed particle-level generation on it. Global quantities are subsequently recomputed from the generated particle clouds and the analysis selections are reapplied.

A deterministic post-generation calibration is applied to restore the conditioned dimuon mass and exact two-body momentum balance. Each event is boosted to the dimuon rest frame, where the muons are treated as on-shell particles. For a conditioned dimuon mass $M$, the target energy and momentum magnitude of each muon are

$$E^* = \frac{M}{2}, \quad p^* = \sqrt{\left(\frac{M}{2}\right)^2 - m_\mu^2}. \tag{4}$$

The direction in the dimuon pair rest frame (PRF) is defined as

$$\hat{\mathbf{n}} = \frac{\mathbf{p}_1^{\mathrm{PRF}} - \mathbf{p}_2^{\mathrm{PRF}}}{|\mathbf{p}_1^{\mathrm{PRF}} - \mathbf{p}_2^{\mathrm{PRF}}|}, \tag{5}$$

and the calibrated four vectors are constructed as

$$p_1^{*'} = (E^*, +p^*\hat{\mathbf{n}}), \qquad p_2^{*'} = (E^*, -p^*\hat{\mathbf{n}}). \tag{6}$$

The calibrated four vectors are then boosted back to the laboratory frame. The calibration magnitude shown in Fig. 3b is defined as the mean relative Euclidean displacement of the muon four-vector components

$$\delta_{\mathrm{cal}} = \left\langle \frac{\| p_i' - p_i \|_2}{\| p_i \|_2} \right\rangle_{i,\mathrm{events}}, \quad \| p_i \|_2 = \sqrt{E_i^2 + p_{x,i}^2 + p_{y,i}^2 + p_{z,i}^2}, \tag{7}$$

where the average is taken over both muons and all generated events.

In the signal region, an XGBoost classifier is trained to distinguish the observed data from the generated background using $\eta_1$, $\eta_2$, $\mathrm{IP3D}_1$, $\mathrm{IP3D}_2$ and $H_T$. The classifier is designed to provide a score with minimal dependence on $m_{\mu\mu}$, thereby limiting artificial sculpting of the dimuon-mass distribution. Fivefold cross-validation is used so that each event is evaluated by a classifier that was not trained on that event. The resulting classifier score $z(\mathbf{x})$ is converted to the signal-to-background likelihood-ratio estimate $\ell(\mathbf{x})$ using equation (2).

Negative values of $\ell(\mathbf{x})$ are clipped to zero, and the resulting non-negative values are used as event weights. The weighted sideband mass spectrum is fitted and interpolated into the signal region to estimate the background yield. A one-sided profile-likelihood test of the background-only hypothesis is then performed using a scaled-Poisson likelihood that accounts for the event weights, and the corresponding local significance is reported.

Four train → test configurations are evaluated: OS → OS, OS → SS, SS → OS and SS → SS. The first label denotes the channel whose sideband data are used to train the conditional background generator and whose signal-region data are used to train the classifier. The second label denotes the channel on which the trained classifier and resonance search are evaluated.

Statistical variability is assessed using eight independently trained generator ensembles. For each ensemble, 200 pseudo-experiments are constructed by resampling 20,000 generated background events and retraining the fivefold classifier with independent randomization. The complete $\ell$-reweighted resonance search is repeated for each pseudo-experiment, giving $8 \times 200 = 1{,}600$ pseudo-experiments per train → test configuration. The reported central value is the median significance and the uncertainties correspond to the central 68% interval of the pooled distribution. The calibration magnitude is summarized separately across the eight independently trained generator ensembles: each ensemble contributes its mean relative four-momentum adjustment, and the displayed centre and error bars are the median and the 16th–84th percentile interval of these eight values.

**Exotic Higgs search.** Events are required to contain at least four jets with $p_T > 30$ GeV and at least three $b$-tagged jets with $p_T > 20$ GeV, with the subleading $b$ jet required to satisfy $p_T > 30$ GeV. The $b$-jet thresholds are defined independently of the inclusive jet requirement. Simulated samples include $H_{SM} \to aa \to 4b$ signal events with $m_a = 30$ GeV and multijet backgrounds with two or four $b$ jets, including $b\bar{b}j$, $b\bar{b}j$ and $b\bar{b}b\bar{b}$ production (Extended Data Table 1).

We compare SPANet[54,55], EVENET–Full, EVENET–SSL and EVENET–Scratch, matched to approximately 20 million trainable parameters in the main model. For pretrained EVENET variants, the backbone learning rate is set to 10% of the base learning rate, and bridge transformer layers between backbone and heads use 30% of the base learning rate. All models are optimized with AdamW for 50 epochs, extended to 100 epochs for training samples below 99,000 events, with early stopping (patience 20). Training subsets are defined as fractions of a reference dataset size of 200,000 events. Event weights are assigned according to the physical process cross-sections, so that the relative contributions of the background samples reflect their expected yields in the combined background pool. These weights are used in both the training loss and the evaluation metrics.

Three task-head configurations are evaluated: Cls, Cls + Asn and Cls + Seg. The classification head predicts two classes: the $H_{SM} \to aa \to 4b$ signal and a single background class combining all multijet processes, while the assignment and segmentation heads serve as auxiliary heads for reconstructing the resonance structure. Search metrics are computed using independent training and evaluation samples for the $m_a$ = 30 GeV signal hypothesis. JES variations are evaluated using 100 pseudo-experiments with nuisance shifts sampled from a Gaussian prior with 1% width. Each trained model is evaluated on the shifted samples without retraining, and shifts in the maximum SIC are recorded relative to the nominal evaluation. Shifts are reported using the normalized responses $\Delta_{SIC} = (\mathrm{SIC} - \langle \mathrm{SIC} \rangle)/\sigma_{SIC}$.

**Top-quark pair quantum-correlation measurement.** In dileptonic top-quark pair production, the spin state of the top-antitop system is described by a two-qubit density matrix $\rho$. In the $(t, \bar{t})$ spin space, $\rho$ is parameterized using the Fano–Bloch decomposition[56,57,82],

$$\rho = \frac{1}{4}(\mathbb{I}_2 \otimes \mathbb{I}_2 + \sum_i B_i^+ \sigma_i \otimes \mathbb{I}_2 + \sum_j B_j^- \mathbb{I}_2 \otimes \sigma_j + \sum_{ij} C_{ij} \sigma_i \otimes \sigma_j), \quad (8)$$

where the coefficients $C_{ij}$ parameterize spin correlations and the polarization terms $B^{\pm}$ are neglected for leading-order LHC top-quark pair production[57]. The coefficients are extracted from angular distributions of decay products in the helicity basis. For analysers A and B with analysing powers $\kappa_A$ and $\kappa_B$, the normalized distribution used for coefficient extraction is[57,60]

$$\frac{1}{\sigma}\frac{d\sigma}{d\cos\theta_{A,i}\cos\theta_{B,j}} = -\frac{1}{2}(1 + \kappa_A \kappa_B C_{ij} \cos\theta_{A,i} \cos\theta_{B,j}) \log|\cos\theta_{A,i} \cos\theta_{B,j}|, \quad (9)$$

Here $\theta_{A,i}$ ($\theta_{B,j}$) denotes the angle between the momentum of the spin analyser A (B) in the top-quark (antitop-quark) rest frame and the corresponding spin quantization axis $i$ ($j$) of the helicity basis. For downstream training, leptons are required to satisfy $p_T$ > 10 GeV and jets $p_T$ > 20 GeV. Inputs include up to four leading-$p_T$ leptons and $b$ jets, as well as up to 6 additional jets reconstructed with a distance parameter $\Delta R = 0.4$. The downstream model uses two task-specific heads: an assignment head that matches reconstructed leptons and $b$ quarks to their parent top-quark decays and a supervised generative head for neutrino reconstruction. For pretrained variants, the backbone learning rate is reduced by a factor of ten relative to task-specific heads, while scratch training uses a uniform learning rate. Early stopping is applied in all configurations. Training sample sizes are expressed relative to a reference size of 1 million events, representative of top-quark analyses using LHC Run 2 data[83]. This reference size defines the training fractions and is distinct from the evaluation sample size.

The evaluation test sample contains 8,195,408 events, of which 445,673 pass the evaluation selection. For evaluation, events are required to contain two $b$ jets and two oppositely charged leptons with $p_T$ > 25 GeV and $|\eta| < 2.5$. The spin-correlation matrix is reconstructed using the decay-angle method, and each $C_{ij}$ component is unfolded in the invariant-mass range (0, 400) GeV following the simulation-based set-up of ref. 60. The unfolded coefficients are combined into $\Delta = |C_{kk} + C_{rr}| - C_{nn} - 1$. Systematic variations are applied at evaluation time using pseudo-experiments: 50 JES variations sampled from a Gaussian prior with 1% width, and 50 soft missing-transverse-energy fluctuations generated from a log-normal distribution with parameters $\mu = 0$ and $\sigma = 1$, truncated to $0 \le \Delta E_T^{miss} \le 5$GeV. The resulting variations are propagated through the full reconstruction and unfolding chain, and shifts in the extracted precision of $\Delta$ are recorded.

## Data availability

The pretraining dataset is publicly available on Hugging Face[84]. Downstream analysis datasets are likewise publicly available: the CMS Open Data hypothesis-grid study[85], the exotic Higgs decay study[86] and the dileptonic top-quark quantum-correlation study[87]. These four Hugging Face datasets are released under the MIT licence and require no special access permissions. The anomaly-detection analysis uses the 2016 CMS Open Data `DoubleMuon` dataset, downloaded directly from the CERN Open Data Portal[45], with event selections described in Methods. CMS Open Data source samples remain subject to the terms specified in their original repository records. CMS Open Data source samples used to construct the hypothesis-grid dataset are listed with their DOIs in Supplementary Information. A curated index of datasets and access instructions is provided at https://evenet-hep.github.io/EveNet-Full/benchmarks/.

## Code availability

The full model implementation for EVENET and all downstream task pipelines is publicly available at https://github.com/EveNet-HEP/EveNet-Full and archived on Zenodo[88]. Pretrained model weights are available at https://huggingface.co/Avencast/EveNet. A curated index of repositories and usage instructions is available at https://evenet-hep.github.io/EveNet-Full/benchmarks/.

## Acknowledgements

We thank R. Gambhir and R. Mastandrea for helpful discussions and constructive feedback on the anomaly-detection analysis. This research used resources of the National Energy Research Scientific Computing Center (NERSC), a US Department of Energy Office of Science User Facility supported by the Office of Science of the US Department of Energy under Contract No. DE-AC02-05CH11231, under NERSC award DDR-ERCAP0034643.

## Author contributions

T.-H.H. designed the network architecture and performed the primary studies, including the large hypothesis-grid analysis, anomaly detection and exotic Higgs decay benchmarks. B.Z. carried out the quantum-correlation studies and related validation. Q.L. developed the initial analysis framework and constructed the Monte Carlo generation and simulation pipeline. Y.-T.C. contributed to the study design and was responsible for drafting and revising portions of the paper. Y.X. contributed through scientific discussions and paper editing. V.M. provided study suggestions and contributed to paper revision. B.N. contributed feedback on methodology and presentation. G.W.-S.H. supervised T.-H.H. S.L. supervised B.Z. and Q.L. S.-C.H. supervised Y.-T.C., Y.X. and Y.Z. Y.Z. conceived of and coordinated the project, developed the training framework and infrastructure, performed the pretraining, and wrote the paper.

## Funding

V.M. discloses support for the research of this work from JST EXPERT-J, Japan (grant number JPMJEX2509). T.-H.H. and G.W.-S.H. disclose support for the research of this work from the National Science and Technology Council (NSTC), Taiwan (grant number 114-2639-M-002-007-ASP). Y.X., Y.-T.C., S.-C.H. and Y.Z. disclose support for the research of this work from the US National Science Foundation (NSF) (grant number PHY-2117997). Y.-T.C. also discloses partial support from the Yushan Fellow Program of the Ministry of Education (MOE), Taiwan (grant number MOE-115-YSFMS-0002-004-P1). Q.L. and B.N. disclose support for the research of this work from the US Department of Energy under Contract No. DE-AC02-76SF00515. B.Z. discloses support for the research of this work from T.D. Lee scholarship. B.Z. and S.L. disclose support for the research of this work from National Key R&D Program of China (grant numbers 2023YFA1606904 and 2023YFA1606900), National Natural Science Foundation of China (grant numbers W2543005 and 12150006) and Shanghai Pilot Program for Basic Research-Shanghai Jiao Tong University (grant number 21TQ1400209).

## Competing interests

The authors declare no competing interests.

## Additional information

**Extended data** is available for this paper at https://doi.org/10.1038/s42256-026-01322-9.

**Supplementary information** The online version contains supplementary material available at https://doi.org/10.1038/s42256-026-01322-9.

**Correspondence and requests for materials** should be addressed to Vinicius Mikuni, Yuan-Tang Chou or Yulei Zhang.

**Peer review information** *Nature Machine Intelligence* thanks Gage Dezoort, and the other, anonymous, reviewer(s) for their contribution to the peer review of this work. Peer reviewer reports are available.

**Extended Data Table 1 | Summary of generated and selected events used for the pretraining and downstream datasets**

| Section | Process | Final states | Generated /M | Sel. Eff. [%] | Used /M | Notes |
|---|---|---|---|---|---|---|
| Pretraining | QCD | inclusive 2,3,4j | 30.4 | 61.9 | 18.9 | |
| Pretraining | $t\bar{t}$ | had. / 1L / 2L + 0,1j | 422 | 16.6 | 70 | |
| Pretraining | $t\bar{t}$W | had. / 1L $t\bar{t}$ + W($\ell v$)/W($qq$) | 692 | 9 | 62.1 | |
| Pretraining | $t\bar{t}$Z | had. / 1L $t\bar{t}$ + Z($\ell\ell$)/Z($qq$) | 630 | 8.9 | 55.8 | |
| Pretraining | $W$+jets | W($\ell v$)/W($qq$) + 0,1,2j | 121 | 39.9 | 48.4 | |
| Pretraining | $Z$+jets | Z($\ell\ell$)/Z($qq$) + 0,1,2j | 129 | 41.2 | 53 | |
| Pretraining | $WW$ | 1L / 2L | 262 | 24.8 | 64.9 | |
| Pretraining | $ZZ$ | 2L / 4L | 196 | 18 | 35.3 | |
| Pretraining | $WZ$ | 1L / 2L / 3L | 256 | 25.8 | 66.1 | |
| Pretraining | $H_{SM}(WW^{*})$ | 0L / 1L / 2L | 192 | 36 | 69.4 | |
| Pretraining | Total | — | 2930.4 | 18.5 | 543 | |
| Downstream (i) Resonant $X \to YH_{SM}$ (CMS Open Data) | $b\bar{b}WW^{*}$ | $b\bar{b} + \ell vqq$ | ~ 0.18 (per mass) | 2–11 | 0.005–0.02 | train:test = 1:1 split per mass point |
| Downstream (i) Resonant $X \to YH_{SM}$ (CMS Open Data) | $t\bar{t}$ | $b\ell v + bqq$ | 144.7 | 2.8 | 4.08 | |
| Downstream (i) Resonant $X \to YH_{SM}$ (CMS Open Data) | Single top | — | 61.54 | 0.15 | 0.09 | |
| Downstream (i) Resonant $X \to YH_{SM}$ (CMS Open Data) | $tW$ | — | 5.04 | 0.79 | 0.04 | |
| Downstream (i) Resonant $X \to YH_{SM}$ (CMS Open Data) | $Z$+heavy flavour | $Z(\ell\ell)$ + b jets | 9.74 | 0.62 | 0.06 | |
| Downstream (ii) Exotic Higgs Decay $H_{SM} \to aa, a \to bb$ | QCD | $bbj$, $bbjj$ | 53.47 | 1.41 | 0.75 | same numbers of events for train and test |
| Downstream (ii) Exotic Higgs Decay $H_{SM} \to aa, a \to bb$ | QCD | $bbbb$ | 1.25 | 13.46 | 0.17 | |
| Downstream (ii) Exotic Higgs Decay $H_{SM} \to aa, a \to bb$ | $H_{SM} \to aa$ | $bbbb$ ($m_a$ = 30) | 7.25 | 1.98 | 0.14 | |
| Downstream (ii) Exotic Higgs Decay $H_{SM} \to aa, a \to bb$ | Total | — | 61.97 | 1.72 | 1.06 | |
| Downstream (iii) Quantum Correlation $t\bar{t} \to 2\ell$ | Train sample | 2L | 12.3 | 12.2 | 1.5 | |
| Downstream (iii) Quantum Correlation $t\bar{t} \to 2\ell$ | Test sample | 2L | 8.2 | 5.49 | 0.45 | |
| Downstream (iii) Quantum Correlation $t\bar{t} \to 2\ell$ | Total | — | 20.5 | — | 1.95 | |
| Downstream (iv) Anomaly Detection (CMS Open Data) | Opposite sign | $m_{\mu\mu} \in [5, 9] \cup [10.6, 16]$ | — | — | 10640 | Absolute data-point counts (not millions) |
| Downstream (iv) Anomaly Detection (CMS Open Data) | Opposite sign | $m_{\mu\mu} \in [9, 10.6]$ | — | — | 1692 | Absolute data-point counts (not millions) |
| Downstream (iv) Anomaly Detection (CMS Open Data) | Same sign | $m_{\mu\mu} \in [5, 9] \cup [10.6, 16]$ | — | — | 6179 | Absolute data-point counts (not millions) |
| Downstream (iv) Anomaly Detection (CMS Open Data) | Same sign | $m_{\mu\mu} \in [9, 10.6]$ | — | — | 989 | Absolute data-point counts (not millions) |
| Downstream (iv) Anomaly Detection (CMS Open Data) | Total | — | — | — | 19500 | Absolute data-point counts (not millions) |

Event counts are reported in millions of events (M), except for the anomaly-detection CMS Open Data channels, which report absolute data-point counts. Columns give the physics process, final-state channels, generated statistics, selection efficiency, and retained statistics after selection. The pretraining corpus comprises ten process categories: QCD, top-quark pairs, top-quark pairs with a W boson, top-quark pairs with a Z boson, W+jets, Z+jets, WW, ZZ, WZ, and Standard Model Higgs-boson production with decay to WW*. Downstream blocks summarise (i) the resonant $X \to YH_{SM}$ CMS Open Data samples used in the hypothesis-grid study, with a 1:1 train/test split per mass point; (ii) the exotic Higgs $H_{SM} \to aa \to 4b$ samples, with equal train and test event counts; (iii) the dileptonic top-quark quantum-correlation samples; and (iv) opposite-sign and same-sign dimuon CMS Open Data yields in the indicated dimuon-mass windows for anomaly detection.

**Extended Data Table 2 | Simulated CMS Open Data samples used in the $X \to YH_{SM} \to b\bar{b}WW^{*}$ analysis**

| Sample | CMS Open Data dataset | $\sigma$ [pb] | $N_{events}$ |
|---|---|---|---|
| tt1l | TTToSemiLeptonic_TuneCP5_13TeV-powheg-pythia8 | 365.35 | $1.45 \times 10^{8}$ |
| DYBJets_100–200 | DYBJetsToLL_M-50_Zpt-100to200_TuneCP5_13TeV-madgraphMLM-pythia8 | 3.222 | $8.85 \times 10^{6}$ |
| DYBJets_200–∞ | DYBJetsToLL_M-50_Zpt-200toInf_TuneCP5_13TeV-madgraphMLM-pythia8 | 0.618 | $8.87 \times 10^{5}$ |
| SingleTop_s | ST_s-channel_4f_leptonDecays_TuneCP5_13TeV-amcatnlo-pythia8 | 3.36 | $5.47 \times 10^{6}$ |
| SingleTop_t | ST_t-channel_top_4f_InclusiveDecays_TuneCP5CR1_13TeV-powheg-madspin-pythia8 | 136 | $2.54 \times 10^{7}$ |
| SingleAntiTop_t | ST_t-channel_antitop_4f_InclusiveDecays_TuneCP5_13TeV-powheg-madspin-pythia8 | 81 | $3.06 \times 10^{7}$ |
| tW_top | ST_tW_top_5f_inclusiveDecays_TuneCP5_13TeV-powheg-pythia8 | 35.8 | $2.49 \times 10^{6}$ |
| tW_antitop | ST_tW_antitop_5f_inclusiveDecays_TuneCP5_13TeV-powheg-pythia8 | 35.8 | $2.55 \times 10^{6}$ |
| Signal | NMSSM_XToYHTo2B2WTo2B2Q1L1Nu* | – | – |

Background samples from the 2016 CMS Open Data Simulation, with dataset names, cross sections $\sigma$ (pb), and generated counts $N_{events}$. Processes include semileptonic $t\bar{t}$, heavy-flavour Drell-Yan in two Z $p_T$ bins, single top (s-/t-channel), and tW. Signal cross sections and yields vary by mass point and are omitted; the full signal list and DOIs are in the Supplementary Information.

## Extended Data Table 3 | Input variables used for EveNet and the XGBoost and TabPFN baseline models

| Model | Feature group | Variables |
|---|---|---|
| EveNet | Particle cloud | $E^i$, $p_T^i$, $\eta^i$, $^i$; $b$-tag and lepton indicators; electric charge $q^i$, for reconstructed object $i$ |
| EveNet | Event-level | $p_T^{miss}$, $^{miss}$; $N_\ell$, $N_b$, $N_j$; $H_T$, $H_T^\ell$; $m(\text{all objects})$; $m(\text{leptons})$; $m(b-\text{taggedjets})$ |
| XGBoost/TabPFN | Event-level | $p_T^{miss}$; number of $b$-tagged jets $N_b$ |
| XGBoost/TabPFN | Lepton ($\ell$) | $p_T^\ell$, $\eta^\ell$; $m_T(\ell, \vec{p}_T^{miss})$ |
| XGBoost/TabPFN | $b$-jet system ($b_1$, $b_2$) | $p_T^{b_i}$, $\eta^{b_i}$; $m_{bb}$; $\Delta R(b_1, b_2)$ |
| XGBoost/TabPFN | Hadronic-$W$ candidate | $p_T^{q_i^h}$, $\eta^{q_i^h}$ $(i = 1, 2)$; $m_{W_{had}} = m(q_1^h q_2^h)$; $p_T^{W_{had}}$; $\Delta R(q_1^h, q_2^h)$ |
| XGBoost/TabPFN | Visible $WW$ system | $m_{WW}^{vis,H} = m(\ell + W_{had})$; $\Delta R(\ell, W_{had})$; $m_T(WW_{had})$ |
| XGBoost/TabPFN | Visible $bbWW$ system | $m_{bbWW}^{vis,h} = m(bb + \ell + W_h)$; $\Delta R(bb, \ell + W_h)$; $m_T(bbWW_h)$ |
| XGBoost/TabPFN | Cross-system correlations | $\lvert\Delta\phi(bb, W_h)\rvert$; $\min\{\Delta R(\ell, q_1^h), \Delta R(\ell, q_2^h)\}$ |
| XGBoost/TabPFN | Topness | $\min\chi^2$; $\Delta R(\ell, b_{lep})$; $m(\ell, b_{lep})$; $m(t_{lep})$; $m(W_{had})$; $\Delta R(q_1, q_2)$; $m(t_{had})$; $\Delta R(W_{had}, b_{had})$ |

EveNet uses particle-cloud kinematic and identification features together with global event-level variables, including missing transverse momentum, object multiplicities, scalar sums of transverse momenta, and invariant masses. XGBoost and TabPFN are trained on an identical set of engineered tabular features. Here $\ell$ denotes the selected charged lepton, $pT^{miss}$ the missing transverse-momentum vector, $b_1$ and $b_2$ the two highest-DeepFlavour b-tagged jets, and $q_1^h$ and $q_2^h$ the light-jet pair forming the hadronic-W candidate under hypothesis h. Two alternative hypotheses assign the two light jets to the hadronic decay in the $H_{SM} \rightarrow WW$ system. In hypothesis (A), the hadronic W candidate is formed from the light-jet pair whose invariant mass is closest to the on-shell W-boson mass, targeting the hadronic W decay to a quark-antiquark pair. In hypothesis (B), the light-jet pair is chosen so that the reconstructed invariant mass of the $\ell\nu q\bar{q}$ system is closest to the Higgs-boson mass. The neutrino longitudinal momentum is obtained by imposing the on-shell W-mass constraint on the leptonic $W \rightarrow \ell\nu$ decay. Kinematic quantities are computed under both hypotheses and included as separate input features. Cluster transverse masses $m_T$ are computed by combining the visible system with the missing transverse-momentum vector. The topness-related variables are reconstructed under the single-lepton top-quark pair hypothesis.

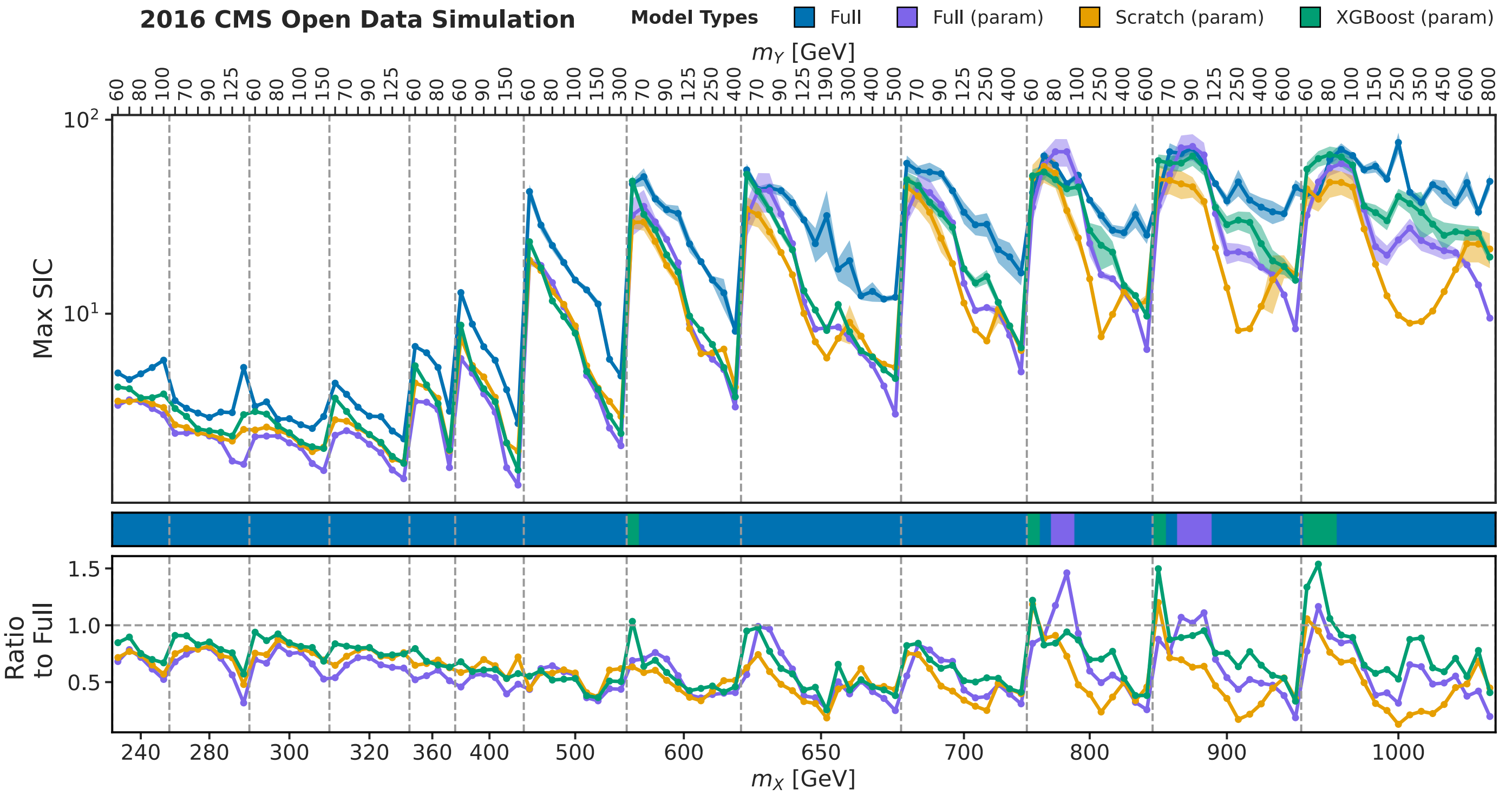

**Extended Data Fig. 1 | Sensitivity across the (mX, mY) grid for X → YH$_{SM}$, with Y→b$\bar{b}$ and HSM → WW$^*$, using parameterised training.** Panels show (i) the maximum significance improvement characteristic (SIC), evaluated with a minimum background yield of 10 events; (ii) a per-bin winner map identifying the method with the highest maximum SIC; and (iii) the ratio of each baseline method's maximum SIC to that of pretrained EveNet-Full. In (i), markers show the central SIC estimates, and shaded bands indicate one-standard-deviation statistical uncertainties obtained by propagating Poisson uncertainties on the signal and background efficiencies into SIC = $\varepsilon S/\sqrt{\varepsilon B}$, where $\varepsilon S$ and $\varepsilon B$ denote the signal and background efficiencies. These uncertainties reflect finite collision-event counts. Signal statistics are identical to those used for individual training and are shown in Fig. 2. Direct comparisons of validation loss are not included.